\documentclass[12pt]{article}
\usepackage{hyperref}
\usepackage{graphicx}
\usepackage{amssymb}
\usepackage{amsmath}
\newcommand{\eps}{\varepsilon}

\numberwithin{equation}{section}

\def\BR{{\mathbb R}}

\def\clf{{\mathcal F}}

\def\sgn{{\rm sgn\ }}
\newtheorem{Pa}{Paper}[section]
\newtheorem{Tm}[Pa]{{\bf Theorem}}
\newtheorem{La}[Pa]{{\bf Lemma}}
\newtheorem{Cy}[Pa]{{\bf Corollary}}
\newtheorem{Rk}[Pa]{{\bf Remark}}

\newtheorem{Ee}[Pa]{{\bf Example}}
\newtheorem{Dn}[Pa]{{\bf Definition}}
\newtheorem{Pn}[Pa]{{\bf Proposition}}

\title{Quantum and classical approaches in statistical physics: some basic inequalities }

\author{Lev Sakhnovich$\footnote{E-mail: lsakhnovich@gmail.com}$}

\date{}

\begin{document}

\maketitle

ORCID: 0000-0003-3590-9583

Courant Institute (retired)

\emph{99 Cove ave. Milford, CT, 06461, USA}

\begin{abstract}
 We present some basic inequalities between the classical and quantum values of free energy, entropy and  mean energy.
We investigate  the transition from the deterministic case
 (classical mechanics)  to the probabilistic case  (quantum mechanics). In the first part of the paper,  we assume that  the reduced Planck constant $\hbar$,  the absolute temperature  $T$, the frequency of an oscillator  $\omega$, and the degree of freedom of a system $N$ are fixed. This approach to the problem of comparing quantum and classical mechanics is new (see \cite{Sakh1}--\cite{Sakh3}).
 
In the second part of the paper, we simultaneously  derive the semiclassical limits for four cases, that is, for $\hbar{\to}0$, $T{\to}\infty$, $\omega{\to}0$, and $N{\to}\infty$.    We note  that only the  case $\hbar{\to}0$ is usually considered in quantum mechanics (see \cite{LL1}). The  cases $T{\to}\infty$ and  $\omega{\to}0$ in quantum mechanics were initially studied   by M. Planck and by A. Einstein, respectively.

\end{abstract}

\section{Introduction}
1. Let us consider some system A in classical mechanics and the same system A in quantum mechanics. The absolute temperature $T$ and the reduced Planck constant $\hbar$ are fixed. There is a general natural  tendency to achieve a minimum of the free energy: $F_{c}(T)$ in classical case and $F_{q}(T,h)$ in quantum  case. We note that
\begin{equation}F_{c}(T)=E_{c}(T)-TS_{c}(T),\quad F_{q}(T,\hbar)=E_{q}(T,\hbar)-TS_{q}(T,\hbar).\label{1.1}\end{equation}
Here, $E_{c}(T)$ and $E_{q}(T,\hbar)$ are mean energies of the system A in classical and in quantum mechanics, respectively, and $S_{c}(T)$ and $S_{q}(T,\hbar)$ are entropies
 of the system A in classical and in quantum mechanics, respectively. The important topic of the classical-quantum duality is discussed (and various
 references are given) in the interesting works \cite{Land, Land1}.

 Recall that the regularized statistical sum $\mathcal{Z}_{r}(T,\hbar)$  is connected with the quantum statistical sum ${Z}_{q}(T,\hbar)$ by the relation
 \begin{equation}\mathcal{Z}_{r}(T,\hbar)=(2{\pi}\hbar)^{N}{Z}_q(T,h),\label{1.2}\end{equation}
where $N$ is  the dimension  of the corresponding coordinate space.
 It is well known that
 \begin{equation}\lim_{\hbar{\to}0}\mathcal{Z}_{r}(T,\hbar)={Z}_c(T),\label{1.3}\end{equation}
The regularized free energy $\mathcal{F}_r(T)$ can be written in the form  \cite{RF, LL2}:
\begin{equation}\mathcal{F}_r(T,\hbar)=-T\log{\mathcal{Z}_r(T,\hbar)}.\label{1.4}\end{equation}
 We introduce the regularized  entropy and mean energy  $\mathcal{S}_{r}$ and $\mathcal{E}_{r}$:
\begin{align}&
\mathcal{S}_{r}(T,\hbar)={S}_{q}(T,\hbar)+N\log (2{\pi}\hbar),\label{1.5}
\\ &
\mathcal{E}_{r}(T,\hbar)={E}_{q}(T,\hbar).\label{1.6}\end{align}
It is easy to see that
\begin{equation}\mathcal{F}_r(T,\hbar)=\mathcal{E}_r(T,\hbar)-T\mathcal{S}_r(T,\hbar). \label{1.7}\end{equation}
The choice of such regularization is explained in section 7.
\begin{Rk}\label{Remark 1.1} Values ${F}_c(T)$ and $\mathcal{F}_r(T,\hbar)$ are also minimal solutions of the corresponding extremal problems (see sections 2 and 3).\end{Rk}

In the first part of the paper, we study the signs of the physical values
\begin{equation} \mathcal{F}_r(T,\hbar)-{F}_c(T),\quad \mathcal{E}_r(T,h)-{E}_c(T),\quad
\mathcal{S}_r(T,\hbar)-{S}_c(T).\label{1.8}\end{equation}
In all the examples, which we investigate in the present paper, the following inequalities hold:
\begin{equation} \mathcal{F}_r(T,\hbar)-{F}_c(T)>0,\quad \mathcal{E}_r(T,\hbar)-{E}_c(T)>0.
\label{1.9}\end{equation}
The expression for entropy satisfies the inequality
\begin{equation} \mathcal{S}_r(T,\hbar)-{S}_c(T)<0
\label{1.10}\end{equation}
in the case of a potential well and
\begin{equation} \mathcal{S}_r(T,\hbar)-{S}_c(T)>0
\label{1.11}\end{equation} in all other cases, which we considered. It follows from \eqref{1.7} that
\begin{equation}\sgn{\mathcal{S}_r(T,\hbar)}=\sgn{[\mathcal{E}_r(T,\hbar)-\mathcal{F}_r(T,\hbar)]}.
\label{1.12}\end{equation}
We note that harmonic oscillators and potential wells being classical objects of research  are also of great current interest
(see, e.g., recent papers \cite{AsK, BaM, Ban, Ges, Kors, Land0} and references therein), and the inequalities we deal with are basic for those cases.

2. There are a number of studies comparing classical and quantum mechanics (see some discussions and references in the recent works
\cite{BoK, H, Kh, Kr, OW, Yan}).
This comparison is made according to different criteria. Our approach is
based on a comparison of average values. We believe
 that such approach is both
natural and important. In this way,
we erase the probability barrier which
separates   classical and quantum  mechanics. (However,  classical mechanics remains deterministic and quantum  mechanics is probabilistic.)
Well-known limit theorems  (e.g., by H. Weyl, A. Pleijel, M. Kac and Kirkwood--Wigner)
are also based on
the comparison of average values.

The extremal principles, for instance,  the Fermat's principle of least time
 \cite{BW} and the principle of least action \cite[Ch. 1]{LL3} remain central in modern physics (see some further references in \cite{DePo, McE, Perl, Sib, Sor, Uzun}).
The named above  extremal principles are  formulated in terms of one physical value.
\begin{Rk}\label{Remark 1.2}  Here, we  investigate the interaction of  three physical values: mean energy E, entropy S and free energy F.  The corresponding extremal principle
(see  sections 2 and 3) may be considered as a cooperative game. In a cooperative game, all players have the same goal.
\end{Rk}

Let's describe the content of the article  in more detail.
At first, we consider the classical case, the quantum case, and the   minimal points $F_{c}(T)$ and $F_{q}(T,\hbar)$ of the corresponding extremal problems.
 In the next step, we compare the results of the two cases (classical and quantum).

From this point of view, we consider
the connection between classical mechanics (deterministic strategy) and quantum mechanics (probabilistic
strategy). We note that such  comparison of the quantum and classical approaches without the requirement  for Planck constant $h$ being small or for absolute temperature
$T$ being large is of essential scientific and methodological interest.

In order to understand  the situation, we study in detail  two following examples: the potential well (polyhedron) and the harmonic oscillator. In the present paper, we (in particular) improve, develop  and generalize our previous results \cite{Sakh1, Sakh2}  and \cite[Ch. 6 and 9]{Sakh3}.

 Using relation \eqref{1.9}, we formulate our hypothesis.\\
\textbf{Hypothesis.} \emph{The members $\mathcal{F}_r(T,\hbar)$ and $\mathcal{E}_r(T,\hbar)$ of the  cooperative game satisfy the equality}
\begin{equation}\sgn[\mathcal{F}_r(T,\hbar)-{F}_c(T)]=
\sgn[\mathcal{E}_r(T,\hbar)-{E}_c(T)].\label{1.13}\end{equation}

 3. In the second part of the paper, we study the semiclassical limits, which is an actively developed domain of research
 (see some recent references in \cite{BaK, Cor0, Cor, Zw}).
  We consider the following four cases.
 $$
\mathrm{Case \, 1:} \quad \hbar{\to}0, \quad \mathrm{Case\, 2:} \quad T{\to}\infty, \quad \mathrm{Case \, 3: }\quad \omega{\to}0, \quad  \mathrm{Case \, 4:} \quad N{\to}\infty. $$
Here, $\omega$ is the frequency of the oscillator and $N$ is the degree of freedom of a system. Usually, the cases 1-3 are investigated by semiclassical analysis  \cite{Mar}
and micro-local analysis
\cite{Land}, and  the case 4 involves the methods of $C^*$-algebras \cite{Land, Sew}.
Our method is based on the fact that we know the eigenvalues of the corresponding boundary problems.
This knowledge allows us to study all the cases 1--4 from one point of view and obtain the corresponding explicit formulas.
\begin{Rk}\label{Remark 1.3} The semiclassical approach in quantum mechanics deals usually with the case $\hbar{\to}0$ (see \cite{LL1}). The first works  dedicated to the
semiclassical approach were written by A. Einstein  $($\cite{Ein}, case $\omega{\to}0)$ and by M. Planck $($\cite{Pl}, case $T{\to}\infty)$.
\end{Rk}

4. As mentioned above, the free energy (in classical mechanics and quantum mechanics, respectively) is considered in sections 2 and 3.
In section 4, we consider in detail  two simple but important examples:  one-dimensional potential well and one-dimensional harmonic oscillator.
These two examples  differ  greatly. However, they  have also some common properties.
 In particular, the following inequalities
 \begin{equation}\mathcal{E}_r(T,\hbar)>E_c(T),\quad \mathcal{F}_r(T,\hbar)>F_c(T)\label{1.14}\end{equation} are fulfilled in both examples (for all $T$ and for $\hbar>0$).
The $N$-dimensional potential well in the domain
\begin{equation}Q_N=\{x:\, x=(x_1, \ldots x_N), \quad 0{\leq}x_k{\leq}a_k \quad (1{\leq}k{\leq}N)\} \label{1.15}\end{equation}
is studied in section 5.
We introduce the corresponding boundary value problem
\begin{align}& -\frac{\hbar^2}{2m}\sum_{k=1}^{N}\frac{\partial^2}{\partial{x_k}^{2}}\psi(x_1,x_2,...,x_N)-
E\psi(x_1,x_2,...,x_N)=0, \label{1.16}
\\ &
\psi|_{\Gamma}=0.\label{1.17}\end{align}
Here, $\Gamma$ stands for the boundary of the domain $Q_N$.
The inequalities \eqref{1.14} are valid in this case too.

Section 6 is dedicated to
the $N$-dimensional oscillator, that is, we consider the differential operator
\begin{equation}\mathcal{L}\psi=-\frac{\hbar^2}{2m}\sum_{k=1}^{N}\frac{\partial^2}{\partial{x_k}^{2}}\psi(x)+
V(x)\psi(x),\quad x{\in}\BR^N, \label{1.18}\end{equation}
where
\begin{equation}V(x)=\sum_{k=1}^{N}\frac{m{\omega_k}{x_k}^2}{2},\quad \omega_{k}>0.\label{1.19}\end{equation}
We  prove the inequalities \eqref{1.14} in the case of the $N$-dimensional oscillator as well.
In section 7, we explain the notions of the regularized statistical sum, mean energy, free energy, and entropy. We note  that the regularized values have the following
important properties:
\begin{align}&\mathcal{Z}_r(T,\hbar){\to}Z_c(T),\quad \mathcal{F}_r(T,\hbar){\to}F_c(T),\quad \hbar{\to}+0,\label{1.20}
\\ &
\mathcal{E}_r(T,h){\to}E_c(T),\quad \mathcal{S}_r(T,\hbar){\to}S_c(T),\quad h{\to}+0.\label{1.21}\end{align}
\begin{Rk}\label{Remark 1.4} The  following facts  are essential for our theory (see section 4).

1. The quotients
 $\mathcal{Z}_r(T,h)/Z_c(T)$ and $\mathcal{E}_r(T,\hbar)/E_c(T)$ depend, in the case of the $1$-dimensional potential well $\,\, 0{\leq}x{\leq}a$,
 on only one variable:
\begin{equation}\mu=h\sqrt{\frac{2\pi}{ma^{2}T}}. \label{1.22}\end{equation}
Thus,  we have
\begin{equation}\mathcal{Z}_r(T,\hbar,a)/Z_c(T,a)=f_1(\mu),\quad \mathcal{E}_r(T,h)/E_c(T)=g_1(\mu).\label{1.23}\end{equation}

2.
The quotients
$\mathcal{Z}_r(T,\hbar,\omega)/Z_c(T,\omega)$ and $\mathcal{E}_r(T,\hbar)/E_c(T)$
depend, in the case of the $1$-dimensional oscillator  $($with frequency $\omega)$,
on  only one variable:
\begin{equation}\tau=\frac{h\omega}{2T}. \label{1.24}\end{equation}
Hence, we have
\begin{equation}\mathcal{Z}_r(T,\hbar,\omega)/Z_c(T,\omega)=f_2(\tau),\quad \mathcal{E}_r(T,\hbar,\omega)/E_c(T,\omega)=g_2(\tau).\label{1.25}\end{equation}

3. The functions $f_1(\mu),\,g_1(\mu)$ and $f_2(\tau),\,g_2(\tau)$ are presented in this paper in an explicit form.
\end{Rk}

In section 8, we study the $N$-dimensional  potential well \eqref{1.15} and  derive the following asymptotic formulas for the statistical sum
\begin{equation}\mathcal{Z}_{r}(T,\hbar)/{Z}_{c}(T)=V_{N}^{-1}[V_N-\rho\frac{V_{N-1}}{2}+...+(-1)^{N}\rho^{N}\frac{V_0}{2^N}]+O(\rho^{N+1}),
\label{1.26}\end{equation}
and for the mean energy
\begin{equation}\mathcal{E}_{r}(T,\hbar)/{E}_{c}(T,\hbar)=1+\rho\frac{V_{N-1}}{2NV_N}+O(\rho^2),
\label{1.27}\end{equation}
where
\begin{equation}\rho=h\sqrt{\frac{\pi}{2mT}}. \label{1.28}\end{equation}
Here, $V_{N}$ is the Lebesgue measure of the domain $Q_{N}$, $V_{N-1}$ is Lebesgue measure of the boundary $\Gamma$, $V_{N-2}$ is the Lebesgue measure of the domain $\Gamma_1$  formed by the intersection of the faces of the domain  $\Gamma$, $\ldots$, and, finally, $V_0$ is the number of the vertices of the polyhedron.

\begin{Ee} \label{Example 1.5} Let $N=3$. Then, $V_3$ is the volume of the polyhedron \eqref{5.1}, $V_2$ is the area of the boundary, $V_1$ is the sum of the lengths of the edges, and $V_0$ is the number of the vertices of the polyhedron \eqref{1.15}.\end{Ee}

In section 9, we  consider  an $N$-dimensional potential well  in a sufficiently general domain
and formulate the well-known results (by H. Weyl, A. Pleijel, and M. Kac) dedicated to  this case.

In section 10, we  consider the general-type potential $V(x)$ without
assuming
that the potential $V(x)$ has the form \eqref{1.19}. We formulate the
well-known Kirkwood--Wigner result (see \cite{Kac1}) dedicated to  the
general-type potential $V(x)$.
 This result implies the assertions

1.The inequality
\begin{equation}
\label{1.29}
\mathcal{F}_r(T,\hbar)>F_c(T) \end{equation}
holds for small $\hbar$ in the general case.

2.The following semiclassical result
\begin{equation}
\label{1.30}
\lim_{\hbar{\to}0}\mathcal{F}_r(T,\hbar)=F_c(T)
\end{equation} is valid.

We consider the Quantum-Classical correspondence at the level of the
inequalities (see \eqref{1.9}-\eqref{1.11}, \eqref{1.14} and \eqref{1.29}).
The traditional semiclassical approach \cite{LL1} is to consider the limit
relations when $\hbar{\to}0$ (see \eqref{1.20}, \eqref{1.21}, and
\eqref{1.30}).

\begin{Rk}\label{Remark{1.6}} We applied   our approach   to Boltzmann equation   \cite[Ch. 10]{Sakh3} and   (jointly with A. Sakhnovich) to Fokker-Planck equation
\cite{ASLS}.\end{Rk}
\section{Free energy (classical case)}
Let us introduce the classical Hamiltonian $H(p,q)$, where $p$ are the corresponding  generalized momenta,
$q$ are the corresponding generalized coordinates. In this case the mean energy $E_{c}$ and the entropy
$S_{c}$ are defined by the formulas
\begin{equation} E_{c}=\int\int H(p,q)P(p,q)dpdq, \label{2.1}\end{equation}
\begin{equation} S_{c}=-\int\int P(p,q)\log{P(p,q)}dpdq, \label{2.2}\end{equation}
\begin{equation} P(p,q){\geq}0,\,\int\int P(p,q)dpdq=1. \label{2.3}\end{equation}
Free energy $F_{c}$ is defined by the formula
\begin{equation} F_{c}=E_{c}-TS_{c} , \label{2.4} \end{equation}
where $T$ is the absolute temperature, $P(p,q)$ is the probability density. In order to find the equilibrium state ($T$ is fixed) we use the
calculus of variations. The corresponding Euler equation has the form
\begin{equation} \frac{\delta}{\delta{P}}[H(p,q)P(p,q)]+TP(p,q)\log{P(p,q)}+\mu{P(p,q)}=0.\label{2.5}\end{equation}
Here, $\frac{\delta}{\delta{P}}$ stands for the functional derivation and $\mu$ is the Lagrange multiplier.
We note that our extremal problem is conditional (see \eqref{2.3}). Formula \eqref{2.5}  yields
\begin{equation} H(p,q)+T+T\log{P(p,q)}+\mu=0.\label{2.6}\end{equation}
From \eqref{2.6} we obtain
\begin{equation}P(p,q)=Ce^{-\lambda{H(p,q)}},\quad \lambda=1/T.\label{2.7}\end{equation}
Formulas \eqref{2.3} and \eqref{2.7} imply that
\begin{equation}P(p,q)=e^{-{\lambda}H(p,q)}/{Z_{c}},
 \label{2.8}\end{equation}
where
\begin{equation}Z_{c}=\int{\int}e^{-{\lambda}H(p,q)}dpdq
 \label{2.9}
 \end{equation}
 is the statistical sum.
 \begin{Rk}\label{Remark 2.1} We deduced above the well-known formulas \eqref{2.8} and \eqref{2.9}.
 These formulas define the equilibrium state. \end{Rk}
 \begin{Rk}\label{Remark 2.2} The inequality
 \begin{equation} \frac{\delta^2}{\delta{P^2}}F_{c}(p,q)=T/P(p,q)>0 \label{2.10}\end{equation}
 shows that the free energy $F_{c}$ has its minimum in the equilibrium state
 which is defined by formulas\eqref{2.8}, \eqref{2.9}.
 \end{Rk}  The strategy of  free entropy $S_{c}$ and mean energy $E_{c}$  is common and is defined by formulas \eqref{2.1}-\eqref{2.3} and \eqref{2.8},\eqref{2.9}.
 \begin{Dn}\label{Definition 2.3}This strategy  defined by formulas \eqref{2.1}-\eqref{2.3} and \eqref{2.8},\eqref{2.9} is optimal.Only by this strategy of all players free energy obtained the minimum.\end{Dn}
 Hence the following statement is valid:
 \begin{Pn}\label{Proposition 2.4}  The entropy $S_{c}$ and the mean energy $E_{c}$ are cooperating members of the game with common goal (minimize free energy)\end{Pn}
 We can write the following relation:
 \begin{equation}{F}_c=\min{F}, \label{2.11}\end{equation}
 where  the value $F$ is defined by formulas \eqref{2.1}-\eqref{2.4} and the free energy ${F}_{c}$ is defined by formulas \eqref{2.1}--\eqref{2.4}
 and  \eqref{2.7}--\eqref{2.9}.
  The free energy ${F}_c(T)$ can be written in the form (see \cite{LL2, RF}):
\begin{equation}{F}_c(T)=-T\log{{Z}_c(T)}.\label{2.12}\end{equation}
  \section{Free energy (quantum  case)}
 Let eigenvalues $E_n$ of the energy operator be given. Consider the mean quantum energy
 \begin{equation} E_q=\sum_{n}P_{n}E_n \label{3.1}\end{equation}
 and the quantum entropy
 \begin{equation} S_q=-\sum_{n}P_{n}\log{P_n}, \label{3.2}\end{equation}
 where $P_n$ are the corresponding probabilities. Hence we have
 \begin{equation}P_n{\geq}0,\quad  \sum_{n=1}^{\infty}P_{n}=1\label{3.3}\end{equation}
Free energy $F_{c}$ is defined by the formula
\begin{equation} F_{q}=E_{q}-TS_{q}, \label{3.4} \end{equation}
where $T>0$ is the absolute temperature.

In order to find the stationary point $P_{st}$ we calculate
\begin{equation}\frac{\partial}{\partial{P_k}}(F_{q}-\mu\sum_{n=1}^{\infty}P_n), \label{3.5}\end{equation}
where $T$ is fixed and $\mu$ is Lagrange multiplier.  We note that our extremal problem is conditional (see \eqref{3.3}). According to \eqref{3.5} we have
\begin{equation} E_n+T+T\log{P_n}-\mu=0. \label{3.6}\end{equation}
It follows from \eqref{3.6} that
\begin{equation} P_n=Ce^{{-\lambda}E_n},\quad \lambda=1/T,\label{3.7}\end{equation}
where $C$ is a constant. Relations \eqref{3.3} imply that
\begin{equation}C=1/{Z_q},\quad Z_q=\sum_{n=1}^{\infty}e^{{-\lambda}E_n},\label{3.8}\end{equation}
where $Z_q$ is the statistical sum.
\begin{Cy}\label{Corollary 3.1}The equilibrium position $P_{st}$ is unique and is defined by the formulas
\eqref{3.7} and \eqref{3.8}.\end{Cy}
By direct calculation we obtain  the equalities:
\begin{equation}\frac{\partial^2}{{\partial}P_{n}^2}F_q=\frac{T}{P_n}>0,\quad
\frac{\partial^2}{{\partial}P_n{\partial}P_k}F_q=0,\quad (n{\ne}k).\label{3.9}\end{equation}
Relations \eqref{3.9} imply the following assertion.
\begin{Cy} \label{corollart 3.2} The equilibrium state $P_{st}$ is a minimum state  of the
free energy $F_q$.
\end{Cy}
The strategy   $P_{n}$
 is common for  entropy $S_{q}$ and   mean energy $E_{q}$.
 Hence, the following statement is  valid.
 \begin{Pn}\label{Proposition 3.3}  The  mean energy $E_{q}$  and the entropy $S_q$ are cooperating members of the game with common goal (minimize the free energy).\end{Pn}
The strategy $P_n$ is optimal (see Definition 2.3)

Using Proposition 3.3, Corollary 3.3 and relations \eqref{1.4}-\eqref{1.6} we obtain the statements:
\begin{Cy} \label{corollary 3.4} The equilibrium state $P_{st}$ is a minimum state  of the regularized free energy $\clf_r$.
\end{Cy}
 \begin{Pn}\label{Proposition 3.5}  The regularized free energy $\mathcal{F}_{r}$, the regularized mean energy $\mathcal{E}_{r}$  and the regularized entropy $S_r$ are cooperating
 members of the game.\end{Pn}
\section{Examples}
In the present section we consider in detail two simple but typical examples.
\begin{Ee}\label{Example 4.1}
\emph{Harmonic oscillator.}\end{Ee}
In the classical case, the Hamiltonian  for the harmonic oscillator has the form:
\begin{equation}H(p,q)=\frac{p^2}{2m}+\frac{m{\omega^2}{q^2}}{2}.\label{4.1}\end{equation}
In the quantum case, the harmonic oscillator  is described  (see \cite[Ch.3]{LL1} and  \cite[Ch. 1]{RF}) by the equation:
\begin{equation}-\frac{\hbar^2}{2m}\frac{d^2}{dx^2}y-(E-\frac{m{\omega^2}{x^2}}{2})y=0,\quad -\infty<x<\infty.
\label{4.2}\end{equation}
The spectrum $E_{n}$ of the boundary problem \eqref{4.1} is defined by the formula:
\begin{equation} E_{n}=\hbar\omega(n-1/2),\quad n=1,2, ...\label{4.3}\end{equation}
Let us consider  ${Z}_c(T)$ and  $\mathcal{Z}_r(T,\hbar)$. It follows from \eqref{2.9} and \eqref{4.3} that
\begin{equation} Z_c(T)=\frac{2{\pi}T}{\omega}. \label{4.4}\end{equation}
Taking into account relations \eqref {1.2}, \eqref{3.8}, and \eqref{4.1}, we have
\begin{equation}\mathcal{Z}_q(T,\hbar)=\frac{1}{2\sinh(\tau)},\quad \mathcal{Z}_r(T,\hbar)=2T\frac{{\pi}\tau}{\omega\sinh(\tau)},\quad \tau=\frac{\hbar\omega}{2T}.\label{4.5}\end{equation}
Hence, the inequality
\begin{equation}\frac{d}{d\tau}[\mathcal{Z}_r(T,\hbar)/{Z}_c(T)]=\frac{\sinh(\tau)-\tau\cosh(\tau)}{\sinh^2(\tau)}<0
\label{4.6}\end{equation}
is valid.
Using relation \eqref{4.6} we obtain the assertion.
\begin{Pn}\label{Proposition 4.2} For harmonic oscillator, the expression  $\mathcal{Z}_r(T,h)/{Z}_c(T)$  monotonically
decreases with respect to $\tau$  and its limit is given by the formula
\begin{equation}\lim_{\tau{\to}0}[\mathcal{Z}_{r}(T,\hbar)/{Z}_c(T)]=1.\label{4.7}\end{equation}
\end{Pn}
 Now,  consider the regularized mean energy $\mathcal{E}_r(T,h)$. It follows from \eqref{2.1} and \eqref{4.3} that
 \begin{equation}{E}_c(T)=T.\label{4.8}\end{equation}
Taking into account relations  \eqref{3.1}, \eqref{3.7},  \eqref{3.8}, \eqref{4.2},
and the  formula
\begin{equation}\sum_{n=1}^{\infty}e^{-an}n=\frac{e^{-a}}{(1-e^{-a})^2},\quad a>0, \label{4.9} \end{equation}
we obtain
\begin{equation}\mathcal{E}_r(T,\hbar)=T\frac{\tau}{\tanh(\tau)},\quad \tau(T,\hbar)=\hbar\omega/(2T).\label{4.10}\end{equation}

The last formula implies that
\begin{equation}\frac{d}{d\tau}[\mathcal{E}_r(T,h)/{E}_c(T)]=[\sinh(2\tau)-2\tau]/(2\sinh^{2}(\tau))>0,\label{4.11}\end{equation}
Relation \eqref{4.11} yields  the next assertion.
\begin{Pn}\label{Proposition 4.3} For harmonic oscillator, the expression  $\mathcal{E}_r(T,h)/\mathcal{E}_c(T)$  monotonically
increases with respect to $\tau$  and
\begin{equation}\lim_{\tau{\to}0}[\mathcal{E}(T,\hbar)/{E}_{c}(T)]=1.\label{4.12}\end{equation}\end{Pn}

Let us turn  to the regularized  free energy $\mathcal{F}_{r}(T,h)$. It follows from  \eqref{2.12}  that
\begin{equation}\mathcal{F}_{r}(T,\hbar)={F}_{c}(T)-T\log[\mathcal{Z}_{r}(T,\hbar)/{Z}_c(T)].\label{4.13}\end{equation}
Taking into account   \eqref{3.4}, we have
\begin{equation}\mathcal{S}_r(T,\hbar)={S}_c(T)+\frac{\tau}{\tanh(\tau)}-1+\log[\mathcal{Z}_r(T,\hbar)/{Z}_c(T)].\label{4.14}\end{equation}
Using relations \eqref{4.5}, \eqref{4.5} and \eqref{4.10}, \eqref{4.11}, \eqref{4.14} we calculate the derivative:
\begin{equation}\frac{d}{d\tau}\mathcal{S}_r(T,\tau)=[\sinh^2(\tau)-\tau^2]/(2\tau\sinh^{2}(\tau))>0.\label{4.15}\end{equation}
We note that
\begin{equation}{F}_c(T)=-T\log{Z}_c(T),\quad {S}_c(T)=1+\log{Z}_c(T),\label{4.16}\end{equation}
where ${Z_c}(T)$ is defined by  \eqref{4.4}.
In view of \eqref{4.14}, we have the proposition below.
\begin{Pn}\label{Proposition 4.4} For harmonic oscillator, the expression   $\mathcal{S}_r(T,\hbar)$  monotonically
increases with respect to $\tau$  and
\begin{equation}\mathcal{S}_r(T,\hbar)>\lim_{\tau{\to}0}\mathcal{S}_r(T,\hbar)={S}_c(T). \quad \label{4.17}\end{equation}\end{Pn}

In view of Propositions 4.2--4.4, and relation \eqref{4.13}, the following corollary is valid.
\begin{Cy}\label{Corollary 4.5} Consider the harmonic  oscillator and let the parameters $T$ and $\omega$ be fixed.
By transition from  deterministic strategy (classical mechanics) to  probabilistic strategy
(quantum mechanics), all members of the  game (i.e., the regularized free energy, the regularized mean energy
and the regularized entropy) increase  when $\tau$ increases. \end{Cy}
 We think  that the comparison of
 the deterministic and probabilistic strategies has a practical and methodological interest.
 Formulas \eqref{4.7}, \eqref{4.12}, \eqref{4.13} and \eqref{4.17} imply the following
 semiclassical limits.

\emph{Case 1}. Let the parameters T and $\omega$ be fixed. If $h{\to}0$ then\\
$$\mathcal{Z}_r(T,\hbar){\to}{Z}_c(T),\quad \mathcal{E}_r(T,\hbar){\to}{E}_c(T),$$
$$\mathcal{F}_r(T,\hbar){\to}{F}_c(T),\quad \mathcal{S}_r(T,\hbar){\to}{S}_c(T).$$

\emph{Case 2}. Let the parameters h and $\omega$ be fixed. If $T{\to}\infty$ then
\begin{equation}\mathcal{Z}_r(T,\hbar){\sim}{Z}_c(T),\quad \mathcal{E}_r(T,\hbar){\sim}{E}_c(T),\quad
\mathcal{S}_r(T,\hbar){\sim}{S}_c(T).
\nonumber\end{equation}

\emph{Case 3}. Let the parameters T and h be fixed. If $\omega{\to}0$ then\\
$$\mathcal{Z}_r(T,\hbar,\omega){\sim}{Z}_c(T,\omega),\quad \mathcal{E}_r(T,\hbar,\omega){\sim}{E}_c(T,\omega),$$
$$\mathcal{F}_r(T,\hbar,\omega){\sim}{F}_c(T,\omega),\quad \mathcal{S}_r(T,\hbar,\omega){\sim}{S}_c(T).$$

For brevity,  we omit $\omega$ in some notations above (when $\omega$ is fixed).

\begin{Rk}\label{Remark 4.6} Recall that the semiclassical approach in quantum mechanics deals usually with the case $(\hbar{\to}0)$ (see \cite{LL1}).\end{Rk}
We need the following result for analytic functions.
The functions $\frac{z}{\sinh(z)}$ and $\frac{z}{\tanh(z)}$ have  Loran series expansions  which converge for all  values   $0<|z|<\pi$:
\begin{align}& \frac{z}{\sinh(z)}=1+\sum_{n=1}^{\infty}\frac{2(1-2^{2n-1})B_{2n}z^{2n}}{(2n)!},\label{A1}
\\ &
\frac{z}{\tanh(z)}=1+\sum_{n=1}^{\infty}\frac{2^{2n}B_{2n}z^{2n}}{(2n)!},\label{A2}\end{align}
where   $B_{2n}$  are  Bernoulli numbers.
\begin{Cy}\label{Corollary 4.7}The functions \begin{equation}f(\tau)=\mathcal{Z}_{r}(T,\hbar)/{Z}_{c}(T,h)=\frac{\tau}{\sinh(\tau)},
\label{A3}\end{equation} and
\begin{equation}g(\tau)=\mathcal{E}_{r}(T,\hbar)/{E}_{c}(T,h)=\frac{\tau}{\tanh(\tau)},
\label{A4}\end{equation}
 are analytic  in the domain $|\tau|<\pi$, and the coefficients of the corresponding Loran series  may be written down explicitly.\end{Cy}
In particular, we have partial expansions:
\begin{equation}f(\tau)=1-B_2{\tau}^2+O(\tau^4),\quad B_2=1/6,\label{A5}\end{equation}
\begin{equation}g(\tau)=1+B_2{\tau}^2+O(\tau^4),\quad B_2=1/6.\label{A6}\end{equation}
 \begin{Ee}\label{Example 4.8}
\bf{Potential well}\end{Ee}
In the classical case the Hamiltonian  for the potential well has the form:
\begin{equation}
H(p,q)=\begin{cases}
\frac{p^2}{2m}, & \text{$q{\in}[0,a];$}\\
+\infty, &\text{otherwise}.\end{cases}\label{4.20}
\end{equation}
In the quantum case the potential well is described \cite[Ch. 3]{LL2} by the equation
\begin{equation}-\frac{\hbar^2}{2m}\frac{d^2}{dx^2}y-Ey=0,\quad y(0)=y(a)=0.
\label{4.18}\end{equation}
The spectrum $E_{n}$ of the boundary problem \eqref{4.18} is given by the formula:
\begin{equation} E_{n}=\frac{\hbar^2\pi^2}{2ma^2}n^2,\quad n=1,2, ...\label{4.19}\end{equation}

It follows from \eqref{2.9} and \eqref{4.20} that
\begin{equation}Z_c(T)=a\int_{-\infty}^{+\infty}\exp(\frac{-p^2}{2mT})dp=a\sqrt{2mT\pi}.\label{4.21}\end{equation}
From \eqref{2.1} and \eqref{4.21} we have
\begin{equation}E_c(T)=a\int_{-\infty}^{+\infty}\frac{p^2}{2m}\exp(\frac{-p^2}{2mT})dp/Z_c(T)=T/2.\label{4.22}\end{equation}
Next, let us consider $\mathcal{Z}_{r}(T,h).$ Using \eqref{1.2}, \eqref{3.8}, and \eqref{4.19}, we derive
\begin{equation}\mathcal{Z}_{r}(T,\hbar)=2\pi{h}\sum_{n=1}^{\infty}e^{-(\pi/4)n^{2}\mu^2},\label{4.24}\end{equation}
where
\begin{equation} \mu=h\sqrt{\frac{2\pi}{ma^{2}T}} \, . \label{4.25}\end{equation}
It follows from \eqref{4.21}, \eqref{4.24}, and \eqref{4.25} that
\begin{equation}\mathcal{Z}_{r}(T,\hbar)/{Z}_{c}(T)=\mu\sum_{n=1}^{\infty}e^{-(\pi/4)n^{2}\mu^2}.\label{4.26}\end{equation}
Hence, we have
\begin{equation}\frac{d}{d\mu}[\mathcal{Z}_{r}(T,\hbar)/{Z}_{c}(T)]=
\sum_{n=1}^{\infty}e^{-(\pi/4)n^{2}\mu^2}(1-(\pi/2)n^2\mu).
\label{4.27}\end{equation}
Relation \eqref{4.27} implies the following assertion.
\begin{La}\label{Lemma 4.9} If the inequality
\begin{equation}\mu{\geq}2/\pi \label{4.28}\end{equation} holds, then
\begin{equation}\frac{d}{d\mu}[\mathcal{Z}_{r}(T,h)/{Z_c}(T)]<0.\label{4.29}\end{equation}\end{La}
Now, let us consider a more difficult case, where
\begin{equation}0<\mu{\leq}2/\pi. \label{4.30}\end{equation}
\begin{La}\label{Lemma 4.10}If the inequality \eqref{4.30}
 holds, then
\begin{equation}\frac{d}{d\mu}[\mathcal{Z}_{r}(T,\hbar)/{Z_c}(T)]<0.\label{4.31}
\end{equation}\end{La}
{\it Proof.} Recall the Poisson formula (see, e.g., \cite{Ev}):
\begin{equation}\sum_{n=0}^{\infty}F(n)=\frac{1}{2}F(0)+\int_{0}^{\infty}F(x)dx+
2\sum_{n=1}^{\infty}\int_{0}^{\infty}F(x)\cos(2{\pi}nx)dx.\label{4.32}\end{equation}
Let us consider the case  $F(x)=e^{-x^{2}(\pi/4)\mu^{2}}$. Since
\begin{equation}\int_{0}^{\infty}\exp(-x^{2}/\nu)\cos(2{\pi}nx)dx=
\frac{\sqrt{\nu\pi}}{2}\exp(-n^2\nu\pi^{2}),\quad n=0,1,2,...\label{4.33}\end{equation}
we have
\begin{equation}\mathcal{Z}_{r}(T,\hbar)/{Z_c}(T)=\mu\{-\frac{1}{2}+\frac{\sqrt{\lambda\pi}}{2}+
\sqrt{\lambda\pi}\sum_{n=1}^{\infty}\exp[-(n\pi)^{2}\lambda]\}, \label{4.34}\end{equation}
where
\begin{equation} \lambda=\frac{4}{\pi\mu^2}.\label{4.35}\end{equation}
Let us calculate the derivative:
\begin{equation}\frac{d}{d\mu}[\mathcal{Z}_{r}(T,\hbar)/{Z_c}(T)]= -1/2+
[16/(\mu^3\pi)]\sum_{n=1}^{\infty}\exp[-(n\pi)^{2}\lambda](n\pi)^{2}. \label{4.36}\end{equation}
We need also the following derivative:
\begin{equation}\frac{d}{d\mu}\{\mu^{-3}\exp[-(n\pi)^{2}\lambda]\}=(8\pi{n^2}\mu^{-6}-3\mu^{-4})\exp[-(n\pi)^{2}\lambda].
\label{4.37}\end{equation}
Taking into account  \eqref{4.30}, we have
\begin{equation}8\pi{n^2}-3\mu^{2}{\geq}8\pi-3\mu^{2}>0.\label{4.38}\end{equation}
It follows from \eqref{4.36}--\eqref{4.38} that  the function $\frac{d}{d\mu}[\mathcal{Z}_{r}(T,\hbar)/{Z_c}(T)]$
monotonically increases and that
\begin{equation}\frac{d}{d\mu}[\mathcal{Z}_{r}(T,\hbar,\mu)/{Z_c}(T)]{\leq}\frac{d}{d\mu}[\mathcal{Z}_{r}(T,\hbar,2/\pi)/Z_{c}(T)]=:G.\label{4.39}\end{equation}
Earlier we omitted for brevity the dependence of the function $\mathcal{Z}_{r}$ on the variable $\mu$.
Formulas \eqref{4.35} and \eqref{4.36}  imply
\begin{equation}G=-1/2+2(\pi)^{4}\sum_{n=1}^{\infty}n^{2}e^{-n^2\pi^3}.\label{4.40}\end{equation}
In order  to estimate   $G$, we study the integral
\begin{equation}\int_{0}^{\infty}x^{2}e^{-x^{2}/\eta}dx=\frac{\eta\sqrt{\eta\pi}}{4}.\label{4.41}\end{equation}
We note that
\begin{equation}\frac{d}{dx}[x^{2}e^{-x^{2}/\eta}]=2xe^{-x^{2}/\eta}(1-x^{2}/\eta).\label{4.42}\end{equation}
Hence, the function
\begin{equation}U(x,\eta)=x^{2}e^{-x^{2}/\eta} \label{4.43}\end{equation}
is monotonically decreasing if $x>\sqrt{\eta}$. In the case under consideration (see \eqref{4.40}), we have
\begin{equation}\eta=\pi^{-3}<1.\label{4.44}\end{equation}
Thus, we  proved that
\begin{equation}\int_{1}^{\infty}U(x,\eta)dx>\sum_{2}^{\infty}U(n,\eta),\quad \eta=\pi^{-3}.\label{4.45}
\end{equation}
Using numerical calculation, we obtain
\begin{equation}\int_{0}^{1}U(x,\eta)dx=0,0025665> U(1,\eta)=3,420(10)^{-14},\quad \eta=\pi^{-3}.\label{4.46}\end{equation}
Then, we have
\begin{equation}\int_{0}^{\infty}U(x,\eta)dx>\sum_{1}^{\infty}U(n,\eta),\quad \eta=\pi^{-3}.\label{4.47}
\end{equation}
According to \eqref{4.41} the equality
\begin{equation}2\pi^{4}\int_{0}^{\infty}U(x,\eta)dx=1/2,\quad \eta=\pi^{-3}\label{4.48}\end{equation} holds. The assertion of the lemma follows
from \eqref{4.40}, \eqref{4.47} and \eqref{4.48}.

Using Lemmas 4.8 and  4.9,  we obtain the proposition below.
\begin{Pn}\label{Proposition 4.11} For the potential well case the expression  $\mathcal{Z}_r(T,\hbar)/{Z_c}(T)$  monotonically
decreases with respect to  $\mu$  and
\begin{equation}\mathcal{Z}_{r}(T,\hbar)/{Z_c}(T)<\lim_{\mu{\to}0}[\mathcal{Z}_{r}(T,\hbar)/{Z_c}(T)]=1.\label{4.49}\end{equation}
\end{Pn}
Let us consider the mean energy $\mathcal{E}_r(T,\hbar)$.
Introduce
\begin{equation}V(T,\hbar):=T\sum_{n=1}^{\infty}e^{(-\pi/4){\mu^2}n^2}(\pi/4){\mu^2}n^2= T\sum_{n=1}^{\infty}e^{-{n^2}/\lambda}({n^2}/\lambda),
\label{4.50}\end{equation}
where $\mu$ and $\lambda$ are defined by the relations \eqref{4.25} and \eqref{4.35}, respectively.
The mean energy $\mathcal{E}_r(T,\hbar)$ may be written in the form
\begin{equation}\mathcal{E}_r(T,\hbar)=V(T,\hbar)/\mathcal{Z}_q(T,\hbar),
\label{4.51}\end{equation}
where $\mathcal{Z}_q(T,\hbar)$ is defined by the relation
\begin{equation}\mathcal{Z}_q(T,\hbar)=\sum_{n=1}^{\infty}e^{-(\pi/4)\mu^2{n^2}}=
\sum_{n=1}^{\infty}e^{-({n^2}/\lambda)}. \label{4.52}\end{equation}
\begin{Rk}\label{Remark 12} The expression $\mathcal{E}_r(T,\hbar)/{E}_c(T)$ depends on only one variable, namely, on  $\mu$ (see \eqref{4.50}--\eqref{4.52}).\end{Rk}

Using relations \eqref{4.50}-\eqref{4.52} we obtain the assertion.
\begin{Pn}\label{Proposition 4.13}If the inequality
\begin{equation}\mu{\geq}2/\pi \label{4.53}\end{equation} holds, then
(in the case of the potential well) we have
\begin{equation}\mathcal{E}_r(T,\hbar)/{E}_c(T)>1.\label{4.54}\end{equation}
\end{Pn}
Let us proceed with a further study of  the case
\begin{equation}\mu{\leq}2/\pi . \label{4.55}\end{equation}
Using  formula \eqref{4.34} and equality
\begin{equation}\mathcal{Z}_q(T,\hbar)=[\mathcal{Z}_r(T,\hbar)/{Z_c}(T)]/\mu, \label{4.56}\end{equation}
we have
\begin{equation}\mathcal{Z}_q(T,\hbar)=-\frac{1}{2}+\frac{\sqrt{\lambda\pi}}{2}+
\sqrt{\lambda\pi}\sum_{n=1}^{\infty}\exp[-(n\pi)^{2}\lambda]. \label{4.57}\end{equation}

Differentiating  \eqref{4.57} with respect to $\lambda$,  we derive
\begin{equation}\frac{d}{d\lambda}\mathcal{Z}_q(T,\hbar)=\frac{1}{4}\sqrt{\frac{\pi}{\lambda}}+
\frac{1}{2}\sqrt{\frac{\pi}{\lambda}}\sum_{n=1}^{\infty}e^{-(n\pi)^{2}{\lambda}}
-\sqrt{\lambda\pi}\sum_{n=1}^{\infty}(n\pi)^{2}e^{-(n\pi)^{2}{\lambda}}\label{4.58}\end{equation}
On the other hand, \eqref{4.52} yields
\begin{equation}\frac{d}{d\lambda}\mathcal{Z}_q(T,\hbar)=\sum_{n=1}^{\infty}e^{-{n^2}/\lambda}\,
({n^2}/{\lambda^2}).\label{4.59}\end{equation}
It follows from \eqref{4.50} and \eqref{4.59} that
\begin{equation}V(T,\hbar)=T\lambda\frac{d}{d\lambda}\mathcal{Z}_q(T,\hbar).\label{4.60}\end{equation}
Let us introduce the functions
\begin{align} & W_{0}(T,\hbar)=1-(\lambda\pi)^{-1/2}+2\sum_{n=1}^{\infty}e^{-(n\pi)^{2}{\lambda}},
\label{4.61}
\\ &
W_{1}(T,\hbar)=1+2\sum_{n=1}^{\infty}e^{-(n\pi)^{2}{\lambda}}-
4\lambda\sum_{n=1}^{\infty}(n\pi)^{2}e^{-(n\pi)^{2}{\lambda}}.\label{4.62}\end{align}
Relations \eqref{4.57},   \eqref{4.58}, and \eqref{4.60}--\eqref{4.62} imply that
\begin{equation}\frac{d}{d\lambda}\mathcal{Z}_q(T,\hbar)=\frac{1}{4}\sqrt{\frac{\pi}{\lambda}}W_{1}(T,\hbar),\quad
\mathcal{Z}_q(T,\hbar)=\frac{\sqrt{\lambda\pi}}{2}W_{0}(T,\hbar).\label{4.63}\end{equation}

It follows from, \eqref{4.51}, \eqref{4.60} and \eqref{4.63} that
\begin{equation}\mathcal{E}_r(T,\hbar)=(T/2)W_{1}(T,\hbar)/W_{0}(T,\hbar).
\label{4.64}\end{equation}

\begin{La}\label{Lemma 4.15}If condition \eqref{4.55} is fulfilled, then
\begin{equation}W_{1}(T,\hbar)>W_{0}(T,\hbar).\label{4.65}\end{equation}\end{La}
\emph{Proof.} Inequality \eqref{4.65} may be rewritten in the equivalent form
\begin{equation} 1>4\lambda^{3/2}\pi^{5/2}\sum_{n=1}^{\infty}n^{2}e^{-(n\pi)^{2}{\lambda}}.
\label{4.66}\end{equation}
It is easy to see that
\begin{equation}{x^2}e^{-\lambda{x^2}{\pi^2}}{\leq}e^{-\lambda{x}{\pi^2}},\quad x{\geq}1,\quad
\lambda{\geq}1.\label{4.67}\end{equation}
It follows from \eqref{4.67} that
\begin{equation} \sum_{n=1}^{\infty}n^{2}e^{-(n\pi)^{2}{\lambda}}{\leq}e^{-\lambda\pi^{2}}/(1-e^{-\lambda\pi^{2}}).
\label{4.68}\end{equation}
Taking into account relations \eqref{4.35} and \eqref{4.55}, we have
\begin{equation}\lambda{\geq}\pi.\label{4.69}\end{equation}
The function $\lambda^{3/2}e^{-\lambda\pi^{2}}$ is monotonically decreasing for $\lambda{\geq}\pi$.
Hence, the inequality
\begin{equation}\lambda^{3/2}e^{-\lambda\pi^{2}}{\leq}2\pi^{3/2}e^{-\pi^{3}},\quad \lambda{\geq}\pi\label{4.70}
\end{equation} holds. Since  $e^{-\pi^{3}}=3,420(10)^{-14}$ (see \eqref{4.46}) the following  relation is valid:
\begin{equation}4\lambda^{3/2}\pi^{5/2}e^{-\lambda\pi^{2}}/(1-e^{-\lambda\pi^{2}})<1
\label{4.71}\end{equation}
The inequality \eqref{4.65} follows from \eqref{4.68} and \eqref{4.71}. The lemma is proved.

Using \eqref{4.64} and \eqref{4.65} we obtain
\begin{Pn}\label{Proposition 4.15}
The following inequality is valid for the potential well in the case $\mu{\leq}2/\pi  :$
\begin{equation}\mathcal{E}_r(T,\hbar)/{E}_c(T)>1.\label{4.72}\end{equation}
\end{Pn}
Formulas \eqref{4.60}--\eqref{4.62} immediately imply the asymptotic relation
\begin{equation}\mathcal{E}_r(T,\hbar)/{E}_c(T)=1/(1-\mu/2)+
O(e^{-4\pi/{\mu^2}}),\quad \mu{\to}+0.\label{4.74}\end{equation}
From propositions 4.2--4.4 and relation \eqref{4.72} we obtain the following theorem.
\begin{Tm}\label{Theorem 4.16} In the case of the potential well, we have
\begin{equation}\mathcal{E}_r(T,\hbar)/{E}_c(T)>\lim_{\mu{\to}+0}[\mathcal{E}_r(T,\hbar)/{E}_c(T)]=1.\label{4.75}\end{equation}
\end{Tm}
 Let us consider the regularized entropy  $\mathcal{S}_{r}(T,\hbar)$ in the case of the  potential well.
It follows from \eqref{1.7} that
\begin{equation}\mathcal{S}_{r}(T,\hbar)=[\mathcal{E}_{r}(T,\hbar)
-\mathcal{F}_{r}(T,\hbar)]/T.
\label{4.76}\end{equation}
According to \eqref{4.56} and \eqref{4.57} the asymptotic equality
\begin{equation}\mathcal{Z}_{r}(T,\hbar)/{Z}_{c}(T,\hbar)=(1-\mu/2)+
O(e^{-4\pi/{\mu^2}}),
\quad \mu{\to}+0\label{4.77}\end{equation} holds.
In view of \eqref{4.74}, \eqref{4.76}, and \eqref{4.77}, we have
\begin{equation}\mathcal{S}_{r}(T,\hbar)=S_{c}(T)-\mu/4+O(\mu^2),
\quad \mu{\to}+0,\label{4.78}\end{equation}
where
\begin{equation}S_{c}(T)=\frac{1}{2}+\log(a\sqrt{2mT\pi}).\label{4.79}
\end{equation}
\begin{Rk}\label{REmark 4.17} The following assertion is valid for the potential well.
By transition from  the deterministic strategy (classical mechanics) to  the probabilistic strategy
(quantum mechanics) two members of the  game (i.e., the regularized free energy and the regularized mean energy) increase.
\end{Rk}
 We can estimate the regularized entropy for potential well only for small $\mu$.
\begin{Rk}\label{Remark 4.18} Formula \eqref{4.77} implies: if $\mu$ is small, then by transition from  the deterministic strategy (classical mechanics) to  the
probabilistic strategy
(quantum mechanics) the regularized entropy for the potential well decreases.
\end{Rk}
Relations \eqref{4.25}, \eqref{4.49}, \eqref{4.14}, \eqref{4.75} and \eqref{4.78} yield the following
 semiclassical limits:

\emph{Case 1}. Let the parameters $T,\, a$ and $m$ be fixed. If $h{\to}0$, then
$$\mathcal{Z}_r(T,\hbar){\to}Z_c(T),\quad \mathcal{E}_r(T,\hbar){\to}E_c(T),$$
$$\mathcal{F}_r(T,\hbar){\to}F_c(T),\quad \mathcal{S}_r(T,\hbar){\to}S_c(T).$$

\emph{Case 2}. Let the parameters $m,\, a$ and $\hbar$ be fixed. If $T{\to}\infty$, then
$$\mathcal{Z}_r(T,\hbar){\sim}Z_c(T),\quad \mathcal{E}_r(T,\hbar){\sim}E_c(T),$$
$$\mathcal{F}_r(T,\hbar){\sim}\mathcal{F}_c(T),\quad \mathcal{S}_r(T,\hbar){\sim}S_c(T).$$

\emph{Case 3}. Let the parameters $\hbar,\, T$ and $m$ be fixed. If $a{\to}\infty$, then
\begin{equation}\mathcal{Z}_r(T,\hbar,a){\sim}Z_c(T,a),\quad \mathcal{E}_r(T,\hbar,a){\sim}{E}_c(T,a),\quad
\mathcal{S}_r(T,\hbar){\sim}{S}_c(T,a).
\nonumber\end{equation}

\emph{Case 4.} Let the parameters $\hbar,\, T$ and $a$ be fixed. If $m{\to}\infty$, then
\begin{equation}\mathcal{Z}_r(T,\hbar,m){\sim}Z_c(T,m),\quad \mathcal{E}_r(T,\hbar,m){\sim}{E}_c(T,m),\quad
\mathcal{S}_r(T,\hbar,m){\sim}{S}_c(T,m).
\nonumber\end{equation}
When the parameters $a$ and $m$  are constant, we omit them for brevity.
\section{$N$-dimensional potential well}
Let us consider the $N$-dimensional potential well $Q_N$:
\begin{equation} 0{\leq}x_k{\leq}a_k, \quad 1{\leq}k{\leq}N.\label{5.1}\end{equation}
We introduce the corresponding boundary problem
\begin{equation}\frac{\hbar^2}{2m}\sum_{k=1}^{N}\frac{\partial^2}{\partial{x_k}^{2}}\psi(x_1,x_2,...,x_N)+
E\psi(x_1,x_2,...,x_N)=0, \label{5.2}\end{equation}
\begin{equation}\psi|_{\Gamma}=0.\label{5.3}\end{equation}
Here, $\Gamma$ stands for the boundary of the domain $Q_N$.
The spectrum of the boundary problem \eqref{5.1}-\eqref{5.3} is given by the formula:
\begin{equation}E_{n_1,...,n_N}=\frac{\hbar^2\pi^2}{2m}\sum_{k=1}^{N}\frac{{n_k}^2}{a_k}
\quad (n_k=1,2,...).\label{5.4}
\end{equation} We need the relation (see \eqref{3.8}):
\begin{equation}-[\frac{\partial}{\partial\beta}{Z}_{q,k}(T,\hbar)]/{Z}_{q,k}(T,\hbar)=
{E}_{q,k}(T,\hbar),\quad \beta=1/T.\label{5.5}\end{equation}
It follows from \eqref{5.4} and \eqref{5.5} that
\begin{equation}{Z}_q(T,\hbar)=\prod_{k=1}^{N}{Z}_{q,k}(T,\hbar),\quad
{E}_q(T,\hbar)=\sum_{k=1}^{N}{E}_{q,k}(T,\hbar),\label{5.6}\end{equation}
where ${Z}_{q,k}(T,\hbar)$ and ${E}_{q,k}(T,\hbar)$ are defined by the relations
\begin{equation}{Z}_{q,k}(T,\hbar)=\sum_{n=1}^{\infty}\exp(-(\pi/4)n^2\mu_{k}^2),
\label{5.7}\end{equation}
\begin{equation}{E}_{q,k}(T,\hbar)=T\sum_{n=1}^{\infty}(\pi/4)n^2\mu_{k}^2
\exp(-(\pi/4)n^2\mu_{k}^2)/{Z}_{q,k}(T,\hbar),
\label{5.8}
\end{equation}
and
\begin{equation} \mu=\hbar\sqrt{\frac{2\pi}{ma_{k}^{2}T}}. \label{5.9}\end{equation}
Let us introduce the regularized statistical sum $\mathcal{Z}_{r}(T,h)$ and the mean energy $\mathcal{E}_{r}(T,\hbar)$ in the $N$-dimensional case:
\begin{equation}\mathcal{Z}_{r}(T,\hbar)=(2{\pi}\hbar)^N{Z}_{q}(T,\hbar),\quad \mathcal{E}_{r}(T,\hbar)={E}_{q}(T,\hbar).\label{5.10}\end{equation}
In the $N$-dimensional case, according to \eqref{4.21} and \eqref{4.22}  we have
\begin{equation}Z_c(T)=(2mT\pi)^{N/2}\prod_{k=1}^{N}a_k,\quad E_c(T)=N(T/2).\label{5.11}
\end{equation}
In view of \eqref{4.77} and \eqref{5.6}, the following asymptotic equality holds:
\begin{equation}\mathcal{Z}_{r}(T,\hbar)/{Z}_{c}(T)=\prod_{k=1}^{N}(1-\mu_{k}/2)+O(e^{-1/\varepsilon^2}),
\quad \varepsilon{\to}0.
\label{5.12}\end{equation}
Here,
\begin{equation}\varepsilon^2= (4/\pi)\max[{\mu_1}^2,{\mu_2}^2,...,{\mu_N}^2].\label{5.13}\end{equation}

Taking into account \eqref{4.74} and \eqref{5.6}, we write:
\begin{equation}\mathcal{E}_r(T,\hbar)=\frac{T}{2}[\sum_{n=1}^{N}(1-\mu_k/2)^{-1}+
O(e^{-1/\varepsilon^2})],  \quad \varepsilon{\to}+0.\label{5.14}\end{equation}
It follows from \eqref{4.78} and \eqref{4.79} that in case of the $N$-dimensional
potential well \eqref{5.1} we have
\begin{equation}\mathcal{S}_{r}(T,\hbar)=S_{c}(T)-\sum_{k}^{N}\mu_{k}/4+O(\varepsilon^2),
\quad \varepsilon{\to}+0,\label{5.15}\end{equation}
where
\begin{equation}S_{c}(T)=\frac{N}{2}+\sum_{k}^{N}\log(a_k\sqrt{2mT\pi}).\label{5.16}
\end{equation}
\begin{Rk}\label{Remark 5.4-} By transition from  the deterministic strategy (classical mechanics) to the
probabilistic strategy
(quantum mechanics), the  regularized entropy decreases for the $N$-dimensional potential well in the case of  small values of $\eps$.
\end{Rk}
We note that  $\varepsilon$ is defined by relation \eqref{5.13}.

\begin{Tm}\label{Theorem 5.2}
The following results are valid for the boundary problem \eqref{5.1}--\eqref{5.3}.

1.The regularized statistical sum  $\mathcal{Z}_r(T,\hbar)$  of a quantum equilibrium system monotonically
decreases with respect to  $\mu_k$  and
\begin{equation}\mathcal{Z}_{r}(T,\hbar)<\lim_{\varepsilon{\to}0}\mathcal{Z}_{r}(T,\hbar)={Z}_c(T)
\label{5.17}\end{equation}

2. We have
\begin{equation}\mathcal{E}_r(T,\hbar)>\lim_{\varepsilon{\to}+0}\mathcal{E}_r(T,\hbar)={E}_c(T)=NT/2,\label{5.18}\end{equation}

3. The regularized free energy  $\mathcal{F}_r(T,\hbar)=-T\log\mathcal{Z}_r(T,\hbar)$  of a quantum equilibrium system monotonically
increases with respect to  $\mu_k$  and
\begin{equation}\mathcal{F}_{r}(T,\hbar)>\lim_{\varepsilon{\to}0}\mathcal{F}_{r}(T,\hbar)={F}_c(T)=-T\log[Z_c(T)].
\label{5.19}\end{equation}\end{Tm}
\begin{Rk}\label{REmark 5.3}The following assertion is valid for the potential well $(N{\geq}1)$.
By transition from  deterministic strategy (classical mechanics) to  probabilistic strategy
(quantum mechanics) two members of the  game (the regularized free energy and the regularized mean energy) increase.
\end{Rk}

\begin{Rk}\label{Remark 5.4}
By transition from  the deterministic strategy (classical mechanics) to  the probabilistic strategy
(quantum mechanics), the regularized entropy for the potential well $(N{\geq}1)$ decreases with respect to $\mu_k$,
if $\varepsilon$ is small.
\end{Rk}
Formulas \eqref{5.12}-\eqref{5.15} imply the following
semiclassical limits.

\emph{Case 1}. Let parameters T, $a_k$ and m be fixed. If $h{\to}0$, then
$$\mathcal{Z}_r(T,\hbar){\to}Z_c(T),\quad \mathcal{E}_r(T,\hbar){\to}E_c(T),$$
$$\mathcal{F}_r(T,\hbar){\to}F_c(T),\quad \mathcal{S}_r(T,\hbar){\to}S_c(T).$$

\emph{Case 2}. Let parameters m, $a_k$ and $\hbar$ be fixed. If $T{\to}\infty$, then
$$\mathcal{Z}_r(T,\hbar){\sim}Z_c(T),\quad \mathcal{E}_r(T,\hbar){\sim}E_c(T),$$
$$\mathcal{F}_r(T,\hbar){\sim}\mathcal{F}_c(T),\quad \mathcal{S}_r(T,\hbar){\sim}S_c(T).$$

\emph{Case 3}. Let parameters $\hbar$, T and m be fixed. If $a_k{\to}\infty \,  (1{\leq}k{\leq}N)$, then
\begin{equation}\mathcal{Z}_r(T,\hbar,a){\sim}Z_c(T,a),\quad \mathcal{E}_r(T,\hbar,a){\sim}{E}_c(T,a),\quad
\mathcal{S}_r(T,\hbar){\sim}{S}_c(T,a).
\nonumber\end{equation}

\emph{Case 4.} Let parameters $\hbar$, T and $a_k$ be fixed. If $m{\to}\infty$, then
\begin{equation}\mathcal{Z}_r(T,\hbar,m){\sim}Z_c(T,m),\quad \mathcal{E}_r(T,\hbar,m){\sim}{E}_c(T,m),\quad
\mathcal{S}_r(T,\hbar,m){\sim}{S}_c(T,m).
\nonumber\end{equation}
When the parameters $a_{k}$ and m  are constant, we omit them for brevity.

\section{$N$-dimensional oscillator}  Schr\"odinger differential operator (quantum case) has the form
\begin{equation}L{\psi}=-\frac{\hbar^2}{2m}\sum_{k=1}^{N}\frac{\partial^2}{\partial{x_k}^{2}}\psi(x)+
V(x)\psi(x),\quad x=[x_1,x_2,...,x_N] ,\label{6.1}\end{equation}
and the corresponding Hamiltonian (classical case) is given by
\begin{equation}H(p,x)=\frac{1}{2m}\sum_{j=1}^{N}p_{j}^{2}+V(x).\label{6.2}\end{equation}
The $N$-dimensional oscillator is determined by the  operator $\mathcal{L}=L$
such that
\begin{equation}V(x)=\sum_{k=1}^{N}\frac{m{\omega_k}^{2}{x_k}^2}{2},\quad \omega_{k}>0.\label{6.3}\end{equation}
The  operator $\mathcal{L}$ has the following spectrum:
\begin{equation}E_{n_1,n_2,...,n_N}(h)=\sum_{k=1}^{N}\hbar\omega_{k}(n_{k}-1/2).\label{6.4}\end{equation}
Using \eqref{4.6} we obtain:
\begin{equation}\mathcal{Z}_{r}(T,h)=\prod_{k=1}^{N}[2T\pi\frac{\tau_k}{\omega_{k}\sinh(\tau_k)}],
\label{6.5}\end{equation}
where
\begin{equation}\tau_{k}=\frac{\hbar\omega_{k}}{2T}.\label{6.6}\end{equation}
Relation \eqref{4.11} implies that
\begin{equation}\mathcal{E}_{r}(T,\hbar)=T\sum_{k=1}^{N}\frac{\tau_k}{\tanh(\tau_k)}.
\label{6.7}\end{equation}
The following corollary is valid.
\begin{Cy}\label{Corollary 6.1}The functions \begin{equation}f(\tau_1,...\tau_N)=\mathcal{Z}_{r}(T,\hbar)/{Z}_{c}(T,\hbar)=\prod_{k=1}^{N}[\frac{\tau_k}{\sinh(\tau_k)}]
\label{6.8}\end{equation} and
\begin{equation}g(\tau_1,...\tau_N)=\mathcal{E}_{r}(T,\hbar)/{E}_{c}(T,\hbar`)=\sum_{k=1}^{N}\frac{\tau_k}{\tanh(\tau_k)}
\label{6.9}\end{equation}
 are analytic  in the domain $|\tau_k|<\pi\,\, (1{\leq}k{\leq}N)$, and the coefficients of the corresponding Loran series expansions may be written in  explicit forms (see \eqref{A1} and \eqref{A2}).\end{Cy}
\begin{Ee}\label{Example 6.2} Partial expansions of $f$ and $g$ are given by the formulas
\begin{equation}f(\tau_1,...\tau_N)=1-B_2\sum_{k=1}^{N}{\tau_{k}}^2+O(\delta^4),\quad B_2=1/6;\label{6.10}\end{equation}
\begin{equation}g(\tau_1,...\tau_N)=1+B_2\sum_{k=1}^{N}{\tau_{k}}^2+O(\delta^4),\quad B_2=1/6,\label{6.11}\end{equation}
where
\begin{equation}\delta=\max\{\tau_1, \tau_2,...,\tau_N\}. \label{6.12+}\end{equation}
\end{Ee}

A number of the results, which we have proved for the case $N=1$ (see section 4, Example 4.1), are valid for the case $N{\geq}1$ as well.  Indeed, taking into account formulas \eqref{6.5}, \eqref{6.7} and \eqref{6.12+}
we obtain the next theorem.
\begin{Tm}\label{Theorem 6.3}
The following results are valid for the system \eqref{6.1}--\eqref{6.3}.\\
1. The regularized statistical sum  $\mathcal{Z}_r(T,\hbar)$  of the corresponding quantum equilibrium system monotonically
decreases with respect to  $\tau_k$  and
\begin{equation}\mathcal{Z}_r(T,\hbar){\sim}Z_c(T)=\prod_{k=1}^{N}\frac{2T\pi}{\omega_k}\quad {\mathrm{for}}\quad
\delta{\to}0.\label{6.13}\end{equation}
2. The regularized mean energy  $\mathcal{E}_r(T,\hbar)$  of the quantum equilibrium system monotonically
increases with respect to  $\tau_k$  and
\begin{equation}\mathcal{E}_{r}(T,\hbar){\sim}{E}_c(T)=TN/2 \quad {\mathrm{for}}\quad \delta{\to}0.
\label{6.14}\end{equation}
3. The regularized free energy  $\mathcal{F}_r(T,\hbar)$  of the quantum equilibrium system monotonically
increases with respect to  $\tau_k$  and
\begin{equation}\mathcal{F}_{r}(T,\hbar){\sim}{F}_c(T)=-T\log[\prod_{k=1}^{N}\frac{2T\pi}{\omega_k}] \quad {\mathrm{for}}\quad \delta{\to}0.
\label{6.15}\end{equation}
4. The regularized entropy  $\mathcal{S}_r(T,\hbar)$  of the quantum equilibrium system monotonically
increases with respect to  $\tau_k$  and
\begin{equation}\mathcal{S}_{r}(T,\hbar){\sim}{S}_c(T) \quad {\mathrm{for}}\quad \delta{\to}0.
\label{6.16}\end{equation}\end{Tm}

\begin{Rk}\label{Remark 6.4}The following assertion is valid for the $N$-dimensional harmonic oscillator $(N{\geq}1)$.
By transition from  the deterministic strategy (classical mechanics) to  the probabilistic strategy
(quantum mechanics) all three  members of the  game (the regularized free energy, the regularized mean energy, and the regularized entropy) increase.
\end{Rk}
Formulas \eqref{6.10}--\eqref{6.12+} imply the following
semiclassical limits.

\emph{Case 1}. Let parameters T and $\omega_k$  be fixed. If $h{\to}0$, then
$$\mathcal{Z}_r(T,\hbar){\to}Z_c(T),\quad \mathcal{E}_r(T,\hbar){\to}E_c(T),$$
$$\mathcal{F}_r(T,\hbar){\to}F_c(T),\quad \mathcal{S}_r(T,\hbar){\to}S_c(T).$$

\emph{Case 2}. Let parameters  h and T be fixed. If $\delta{\to}0$, then
$$\mathcal{Z}_r(T,\hbar){\sim}Z_c(T),\quad \mathcal{E}_r(T,\hbar){\sim}E_c(T),$$
$$\mathcal{F}_r(T,\hbar){\sim}F_c(T)\quad \mathcal{S}_r(T,\hbar){\sim}S_c(T).$$

\emph{Case 3}. Let parameters $\hbar$ and $\omega_k$  be fixed. If $T{\to}\infty$, then
\begin{equation}\mathcal{Z}_r(T,\hbar,a){\sim}Z_c(T,a),\quad \mathcal{E}_r(T,\hbar,a){\sim}{E}_c(T,a),\quad
\mathcal{S}_r(T,\hbar){\sim}{S}_c(T,a).
\nonumber\end{equation}

When the parameters $\omega_{k}$   are constant, we omit them for brevity.

\section{Regularization}
Let us recall the notion of the regularization. The volume $V$ of the phase space is defined (see, e.g., \cite{LL2}) by the relation
\begin{equation} dV=\frac{dpdq}{(2\hbar\pi)^N}.\label{7.1}\end{equation}
Hence, in view of \eqref{2.9} we have
\begin{equation}\mathcal{Z}_r(T,\hbar)=(2h\pi)^{N}\mathcal{Z}_q(T,\hbar).\label{7.2}\end{equation}
Using \eqref{2.12} we write
\begin{equation}\mathcal{F}_r(T,\hbar)=-T\log[\mathcal{Z}_r(T,\hbar)].\label{7.3}\end{equation}
According to \eqref{2.1} and \eqref{2.3}, the following equality is valid:
\begin{equation}\mathcal{E}_{c}(T)=\int\int H(p,q)\tilde{P}(p,q)dpdq\, \Big/ \, \int\int \tilde{P}(p,q)dpdq,\label{7.4}\end{equation}
where $\tilde{P}(p,q){\geq}0.$ It follows from \eqref{7.1} and \eqref{7.4} that
\begin{equation}\mathcal{E}_{r}(T,\hbar)={E}_{q}(T,\hbar). \label{7.5}\end{equation}
Relations \eqref{2.4} and \eqref{3.4} imply that
\begin{equation}\mathcal{S}_r(T,\hbar)=[\mathcal{E}_r(T,\hbar)-\mathcal{F}_r(T,\hbar)]/T.\label{7.6}\end{equation}
\section{Geometrical interpretation}.
Formula \eqref{5.12} may be rewritten as
\begin{equation}\mathcal{Z}_{r}(T,\hbar)/{Z}_{c}(T)=U_{N}^{-1}[U_N-U_{N-1}\rho+...+(-1)^{N}\rho^{N}U_0]+O(e^{-1/\varepsilon^2}),
\label{8.1}\end{equation}
where $\varepsilon{\to}0$,
\begin{equation}\rho=\hbar\sqrt{\frac{\pi}{2mT}}, \label{8.2}\end{equation}
and
\begin{equation}U_k=\sum {a_{i_1}...a_{i_k}},\quad 1{\leq}i_1<i_2<...<i_k=N,\quad U_0=1. \label{8.3}\end{equation}
In view of \eqref{8.3}, the coefficients $U_k$ have clear geometrical interpretation, namely
\begin{equation}U_k=V_k/2^{N-k},\quad 1{\leq}k{\leq}N, \label{8.4}\end{equation}
Here, $V_{N}$ is Lebesgue measure of the domain $Q_{N}$, $V_{N-1}$ is Lebesgue measure of its boundary $\Gamma$, $V_{N-2}$ is Lebesgue measure of the domain $\Gamma_1$  formed by the intersection of the faces of the domain  $\Gamma$, $\ldots$, and, finally, $V_0$ is the number of the vertices of the polyhedron.

\begin{Ee}\label{Example 8.1} Let $N=3$. Then, $V_3$ is the volume of the polyhedron \eqref{5.1}, $V_2$ is the area of the boundary, $V_1$ is the sum of the lengths of the edges and $V_0$ is the number of the vertices of the polyhedron \eqref{5.1}.\end{Ee}
Relations   \eqref{8.1} and \eqref{8.4} imply (see \cite{Sakh2}) that
\begin{equation}\mathcal{Z}_{r}(T,h)/{Z}_{c}(T)=V_{N}^{-1}[V_N-\rho\frac{V_{N-1}}{2}+...+(-1)^{N}\rho^{N}\frac{V_0}{2^N}]+O(e^{-1/\varepsilon^2}),
\label{8.5}\end{equation}
where $\varepsilon{\to}0$.
Formula \eqref{5.11} yields
\begin{equation}Z_c(T)=(2mT\pi)^{N/2}V_N.\label{8.6}\end{equation}

It follows from \eqref{5.14} and \eqref{5.18} that the mean energy $\mathcal{E}_{r}(T,\hbar)$ satisfies the relation
\begin{equation}\mathcal{E}_{r}(T,\hbar)/{E}_{c}(T,\hbar)=1+\rho\frac{V_{N-1}}{2NV_N}+O(\rho^2).\label{8.7}\end{equation}
\begin{Cy}\label{Corollary 8.2} Let the potential well $Q_N$ be  defined by \eqref{5.1}.
According to  \eqref{5.13} and \eqref{8.2},  formulas \eqref{8.5} and \eqref{8.7} are valid
in the following cases:\\
\emph{Case 1}. Parameters T and m are fixed, $\hbar{\to}0$.\\
\emph{Case 2}. Parameters m and  $\hbar$  are fixed,  $T{\to}\infty$.\\
\emph{Case 3}. Parameters T and $\hbar$  are fixed,  $m{\to}\infty$.\end{Cy}
\textbf{Hypothesis.} \emph{For a bounded  convex  non-degenerate  $N$-dimensional polyhedron
$Q_N$, $N{\geq}2$ we have }
\begin{equation}\mathcal{Z}_{q}(T,1)=(4\pi{t})^{-N/2}[V_N-\rho\frac{V_{N-1}}{2}+\sum_{k=2}^{N}(-1)^{k}(\rho/2)^{k}A_k
+O(\rho^{N+1})],
\label{8.8}\end{equation}
\emph{where $\rho=\sqrt{{\pi}t}\,{\to}\,0$, $\hbar=1$, $m=1/2$, $t=1/T$ and}
\begin{equation}A_k=\sum_{j=1}^{M_{N-k}}[\pi/\omega_{N-k}(j)-\omega_{N-k}(j)/\pi](3/2)V_{N-k}(j).\label{8.9}\end{equation}
Here, $V_{N}$  is the volume of $Q_N$ and $V_{N-1}$ is the measure of its boundary (similar to the case \eqref{5.1}),
 $M_{N-k}$ is the number of the $(N-k)$-dimensional faces $F_{N-k}(j)$ of $Q_N$,
$V_{N-k}(j)$ is the (N-k)-dimensional measure of the face $F_{N-k}(j)$, and $\omega_{N-k}(j)$ is the mean magnitude of the
$(N-k+1)$-dimensional dihedral angles at the face
$F_{N-k}(j)$.

\begin{Rk} \label{Remark 8.3}
The vertices of the polyhedron  (in the hypothesis above) have  the dimension $0$ and the measure $1$. \end{Rk}
\textbf{Support of the hypothesis.}\\
1. If  $Q_N$ is given by \eqref{5.1}, then $\omega_{N-k}(j)=\pi/2$. In this case, the formulas \eqref{8.8} and \eqref{8.9} are valid (compare with \eqref{8.5} and \eqref{8.6}).\\
2. If $N=2$, then the formulas \eqref{8.8} and \eqref{8.9} are valid as well (see  the result of M.~Kac in \cite{Kac1} and the work \cite{MS}).\\
3. If $N{\geq}3$, then the coefficient $A_2$ in \eqref{8.8} has the form \eqref{8.9}  (see the papers \cite{Fed, Gol} by B.V. Fedosov and by A. Goldman and P. Calka).

\section{$N$-dimensional potential well, \\ historical remarks}
Consider the Schr\"odinger differential operator
\begin{equation}L{\psi}=-\frac{\hbar^2}{2m}\sum_{k=1}^{N}\frac{\partial^2}{\partial{x_k}^{2}}\psi(x), \quad x{\in}Q_{N},
 \label{9.1}\end{equation}where $Q_{N}$ is  a bounded domain with a piecewise-smooth boundary $\Gamma$. Introduce the following boundary condition:
 \begin{equation}\psi|_{\Gamma}=0.\label{9.2}\end{equation}
H. Weyl proved in 1911 \cite{WH} that
\begin{equation}\mathcal{Z}_r(T,\hbar){\sim}(2mT\pi)^{N/2}V_{N},\quad \rho{\to}+0,\label{9.3}\end{equation} where $V_{N}$ is Lebesgue measure of the domain $Q_{N}$.
For the case $N=2$, A.~Pleijel proved \cite{PA} the relation:
\begin{equation}\mathcal{Z}_r(T,\hbar){\sim}(2mT\pi)^{1/2}(V_{2}-\rho\frac{V_{1}}{2}),\quad \rho{\to}+0,\label{9.4}\end{equation} where $V_2$ is the area of the region $Q_2$, $V_{1}$ is the length  of the boundary $\Gamma$, and $\rho$ is defined by the relation \eqref{8.2}.
The case  $N{\geq}2$  was investigated in the papers \cite{Kac1, MS, WH}.
For a smooth boundary $\Gamma$, the following result was obtained:
\begin{equation}\mathcal{Z}_r(T,h){\sim}(2mT\pi)^{N/2}(V_{N}-\rho\frac{V_{N-1}}{2}),\quad \rho{\to}+0.\label{9.5}\end{equation}

\section{Semiclassical limit, \\ Kirkwood-Wigner expansion}
In the present section, we  study the case when $h$ is small $(\hbar{\to}0)$.
Let us consider again the Schr\"{o}dinger differential operator (quantum case)
\begin{equation}L{\psi}=-\frac{\hbar^2}{2m}\sum_{k=1}^{N}\frac{\partial^2}{\partial{x_k}^{2}}\psi(x)+V(x)\psi(x), \quad x{\in}\BR^{N},
 \label{10.1}\end{equation}
and the corresponding Hamiltonian (classical case)
\begin{equation}H(p,x)=\frac{1}{2m}\sum_{k=1}^{N}p_{k}^{2}+V(x).\label{10.2}\end{equation}
 J.G. Kirkwood and E. Wigner obtained the following result:
\begin{Tm} \label{Theorem 10.1} Let the inequalities
\begin{align}& Z_{0}(T)=\int{e^{-V(x)/T}}dx<\infty, \label{10.3}
\\ &
Z_{2}(T)=\frac{1}{24{m}T^3}\int{e^{-V(x)/T}}\|\mathrm{grad}\,[V(x)]\|^{2}dx<\infty \label{10.4}\end{align}
be valid.  Then, the relations
\begin{align}&\mathcal{Z}_{r}(T,\hbar)=(2h\pi)^{N}Z_{q}(T,\hbar)=(2{\pi}mT)^{N/2}[Z_{0}(T)-\hbar^{2}Z_{2}(T)+O(\hbar^4)],\nonumber
\\ &
{Z_c}(T)=(2{\pi}mT)^{N/2}Z_{0}(T)\nonumber
\end{align}hold.
\end{Tm}
From Theorem 10.1 and relation \eqref{7.3} we obtain:
\begin{Cy}\label{Corollary 10.2} Let the conditions \eqref{10.3} and  \eqref{10.4} be fulfilled.
Then,
\begin{equation} \mathcal{F}_{r}(T,\hbar)=F_{c}(T)+h^{2}TZ_{2}(T)/Z_{0}(T)+O(\hbar^4),\quad
\label{10.7}\end{equation}\
where
\begin{equation}F_{c}(T)=-T\log{Z_{c}(T)} .\label{10.8}\end{equation}\end{Cy}
Formula \eqref{10.7} is derived in the book \cite[Ch 3, section 33]{LL2}.
\begin{Cy}\label{Corollary 10.3} Let the conditions of Theorem 10.1 be fulfilled. Then,
 the regularized free energy  $\mathcal{F}_r(T,\hbar)$  of the quantum equilibrium system satisfies the relation
\begin{equation}\mathcal{F}_{r}(T,\hbar)>\lim_{\hbar{\to}0}\mathcal{F}_{r}(T,\hbar)=\mathcal{F}_c(T).
\label{10.9}\end{equation}\end{Cy}
Let us compare Theorem 10.1 and Corollary 10.2
 with the results of section~6 for the $N$-dimensional oscillator.
It follows from \eqref{6.9}--\eqref{6.11} that in the case of the $N$-dimensional oscillator we have
\begin{align}& \mathcal{Z}_{r}(T,\hbar)=\prod_{k=1}^{N}[(2T\pi)/{\omega_k}[1-\sum_{k=1}^{N}\frac{(\hbar\omega_k)^2}{24T^2}+O(h^4)],\label{10.10}
\\ &
\mathcal{E}_{r}(T,\hbar)=T\sum_{k=1}^{N}[1+\frac{(\hbar\omega_k)^2}{12T^2}+O(\hbar^4)].\label{10.11}\end{align}
Using \eqref{7.3}, \eqref{7.6} and \eqref{10.10}, \eqref{10.11}, we derive
\begin{align}& \mathcal{F}_{r}(T,\hbar)=F_{c}(T)+\hbar^{2}T
\sum_{k=1}^{N}\frac{(\hbar\omega_k)^2}{24T^2}+O(\hbar^4)],\label{10.12}
\\ &
\mathcal{S}_{r}(T,\hbar)=S_{c}+\sum_{k=1}^{N}\frac{(\hbar\omega_k)^2}{24T^2}+O(\hbar^4)].\label{10.13}\end{align}
By comparing relations \eqref{10.10} and \eqref{10.12} we obtain the following assertion.
\begin{Cy}\label{Corollary 10.4} In the case of the $N$-dimensional oscillator, we have:
\begin{equation}Z_{2}(T)/Z_{0}(T)=\sum_{k=1}^{N}\frac{(\omega_k)^2}{24T^2}.\label{10.14}\end{equation}
\end{Cy}
\begin{Cy}\label{Corollary 10.5}In the case of the $N$-dimensional oscillator, formulas \eqref{10.11} and \eqref{10.13} may be rewritten in the form:
\begin{align}&\mathcal{E}_{r}(T,\hbar)=E_{c}(T)+2\hbar^{2}TZ_{2}(T)/Z_{0}(T)+O(\hbar^4),\label{10.15}
\\ &
\mathcal{S}_{r}(T,\hbar)=S_{c}(T)+\hbar^{2}Z_{2}(T)/Z_{0}(T)+O(\hbar^4)\label{10.16}.\end{align}
\end{Cy}

\section{Appendix. The limit $N{\to}\infty$}
 The general situation in the case $N{\to}\infty$ is discussed in the interesting work
\cite{Land} by N.P. Landsman. Numerous useful references on this topic are given in \cite{Land}
as well. In this section, we consider two simple examples: the $N$-dimensional potential well and the
$N$-dimensional oscillator. For these examples we obtain concrete formulas, which show the interconnections between quantum and classical results.

{\it We assume that  $N\to \infty$ together with  $\varepsilon_{N}{\to}0$ in \eqref{11.3} and \eqref{11.4}. Similarly, we assume that $N\to \infty$ together with $\delta_{N}{\to}0$
in \eqref{11.8} and \eqref{11.9}.}
\begin{Ee}\label{Example 11.1} The $N$-dimensional potential well.\end{Ee}
Let  the relations \eqref{5.1}--\eqref{5.3} be fulfilled. Introduce $\varepsilon_N>0$ and $\nu_N>0$ by the equalities
\begin{equation}{\varepsilon_N}^2= \max\{{\mu_1}^2,{\mu_2}^2,...,{\mu_N}^2\}\label{11.1}\end{equation}
and
\begin{equation}{\nu_N}^2= \min\{{\mu_1}^2,{\mu_2}^2,...,{\mu_N}^2\}.\label{11.2}\end{equation}

Using \eqref{5.12}, we derive:
\begin{equation}\mathcal{E}_r(T,\hbar,N)/{E}_c(T,N)=1+\frac{1}{N}\sum_{k=1}^{N}\mu_k +O({\varepsilon_N}^2),\quad \varepsilon_{N}{\to}0,\label{11.3}\end{equation}
\begin{equation}\mathcal{Z}_r(T,\hbar,N)/{Z}_c(T,N)=1-\frac{1}{N}\sum_{k=1}^{N}\mu_k +O(\varepsilon_{N}^2),\quad \varepsilon_{N}{\to}0.\label{11.4}\end{equation}
The parameters $\mu_k$ are defined by \eqref{5.9} and the following inequalities hold:
\begin{equation}\nu_N{\leq}\frac{1}{N}\sum_{k=1}^{N}\mu_k{\leq}\varepsilon_N.\label{11.5}\end{equation}
\begin{Ee}\label{Example 11.2} The $N$-dimensional oscillator.\end{Ee}
We assume that the relations \eqref{6.1} and \eqref{6.3} are valid. We introduce $\delta_N$ and $\varkappa_N$ by the equalities
\begin{equation}\delta_N=\max\{\tau_1, \tau_2,...,\tau_N\} \label{11.6}\end{equation}
and
\begin{equation}\varkappa_N=\min\{\tau_1, \tau_2,...,\tau_N\}. \label{11.7}\end{equation}
Using
(6.8) and (6.9), we obtain:
\begin{equation}\mathcal{E}_r(T,\hbar,N)/{E}_c(T,N){=}1+\frac{1}{6N}\sum_{k=1}^{N}{\tau_k}^2+O({\delta_N}^4),\quad \delta_N{\to}0,\label{11.8}\end{equation}
\begin{equation}\mathcal{Z}_r(T,h,N)/{Z}_c(T,N){=}1-\frac{1}{6N}\sum_{k=1}^{N}{\tau_k}^2+O({\delta_N}^4),\quad \delta_N{\to}0.\label{11.9}\end{equation}
Here, the parameters $\tau_k$ are defined in \eqref{6.6}.  We note that
\begin{equation}{\varkappa_N}^2{\leq}\frac{1}{N}\sum_{k=1}^{N}{\tau_k}^2{\leq}{\delta_N}^2.\label{11.10}\end{equation}
\begin{Rk} Formulas \eqref{11.3}, \eqref{11.4}, \eqref{11.8}, and \eqref{11.9} are also valid for the case of the fixed $N$.
\end{Rk}

\textbf{Acknowledgements}
The author is very grateful to A. Sakhnovich and I. Tydniouk for discussions and essential remarks.

\end{document}